\documentclass[%
 reprint,
 amsmath,amssymb,
 aps,
 prl
]{revtex4-1}

\usepackage{graphicx}
\usepackage{dcolumn}
\usepackage{bm}
\usepackage{hyperref}

\usepackage{epstopdf}
\begin{document}


\title{Intrinsic long range antiferromagnetic coupling in dilutely V doped CuInTe2}


\author{Weiyi Gong}
\author{Ching-Him Leung}
\author{Chuen-Keung Sin}
\author{Jingzhao Zhang}
\author{Xiaodong Zhang}
\author{Bin Xi}
\author{Junyi Zhu}
\email{jyzhu@phy.cuhk.edu.hk}
\affiliation{
 Department of Physics, The Chinese University of Hong Kong, Hong Kong SAR, China
}

\begin{abstract}
Despite the various magnetic orders mediated by superexchange mechanism, the existence of a long range antiferromagnetic (AFM) coupling is unknown. Based on DFT calculations, we discovered an intrinsic long range AFM coupling in V doped CuInTe$_2$. The AFM coupling is mainly due to the $p-d$ coupling and electron redistribution along the interacting chains. The relatively small energy difference between $p$ and $d$ orbitals and the small energy difference between d orbitals of the dopants and that of stepping stone sites can enhance the stability of this AFM configuration. A multi-bands Hubbard model was proposed to provide fundamental understanding to the long range AFM coupling in chalcopyrite diluted magnetic semiconductors(DMS).
\end{abstract}

\maketitle

\section{\label{introduction}Introduction}
Anti-ferromagnetic(AFM) order is essential in many spintronic devices such as tunnel junctions, spin valves, Hall devices, AFM/FM bilayers.\cite{Wang:2017hn,2018RvMP...90a5005B,2016NatMa..15..535F}. Compared to ferromagnets, AFM materials have high information storage densities, high reading and writing speed, low power consumption, and tetra-hertz spin oscillation frequencies\cite{2014PhRvL.113e7601C}. AFM order has been observed in various metallic alloys\cite{Zajac:2001dq}, in which magnetism tuning can be challenging. On the other hand, AFM order in diluted magnetic semiconductors(DMS) might be relatively easy to tune because prototypical semiconductor properties are preserved. 

In DMS, a relatively low concentration of transition metal atoms are doped, compared to magnetic semiconductors. This unique property makes tuning of magnetic order in DMS easier than typical magnetic semiconductors. In early theoretical investigations, it has been found that the magnetic moment of magnetic dopants tend to form antiparallel coupling in large magnetic doping concentrations and small simulation cells in conventional semiconductors such as V doped GaAs under carrier free condition\cite{2007JPCM...19Q6227B}. However, such a coupling may not exist in DMS due to the limited dopant concentrations. Whether a long range AFM coupling exists or not in DMS is still an open question. 

Generally speaking, superexchange\cite{1955PhRv..100..564G} and coupling between localized spins and carriers\cite{1954PhRv...96...99R,1956PThPh..16...45K,1957PhRv..106..893Y} are the major mechanisms to induce an AFM coupling in materials. Here, the magnetic configurations are often in the type of ABA chains, where A sites are magnetic elements. In DMS, the short range AFM coupling leads to an intrinsic difficulty to realize long range AFM order. Therefore, to realize the long range AFM order, there must be other mechanisms involved. Recently, topological surface states\cite{2009NatPh...5..438Z} and stepping stone mechanism have been proposed to explain the ferromagnetic coupling in Cr doped (Sb, Bi)$_2$Te$_3$\cite{2017arXiv170104943C}. Stepping stone mechanism is based on spin polarized s lone pair states of cations in C sites of spin chains of ABCBA type. Whether stepping stone mechanism can be extended to spin polarized d states that may stabilize the long range AFM coupling is still unknown, except one report on a possible long range AFM coupling in Mn doped LiZnAs\cite{2018arXiv180301179Z}, which is relatively unstable. Also, no quantitative model has been constructed to describe the stepping stone mechanism and fundamental physical understanding is poor.  

In this letter we proposed a new kind of long range AFM coupling of magnetic dopants in transition metal(TM) chalcopyrite, CuInTe$_2$, which is an intrinsic narrow band gap semiconductor. The long range is defined as ABCBA type or longer spin chains. An extended super-exchange model was constructed to explain the underlying mechanism which gives rise to the intrinsic AFM coupling in DMS. This microscopic mechanism may lead to the realization of macroscopic AFM order, which can be checked by large scale simulations that is beyond the capability of DFT.

\section{\label{method}Methods}
All calculations were performed using projected augmented plane wave method\cite{1994PhRvB..5017953B} and density functional theory with Perdew-Burke-Ernzerhof generalized gradient approximation(GGA) \cite{2009PhRvL.102c9902P} as implemented in VASP code\cite{1996PhRvB..5411169K}. Atoms were relaxed with force tolerance of 0.01 eV\AA$^{-1}$. A plane wave energy cut-off of 300 eV was used in all calculations. The Brillouin zone integrations were performed by using $\Gamma$ centered  $5\times5\times5$ k-points grid. A strong correlation effect was considered for transition metal, and DFT+$U$ method was used\cite{1998PhRvB..57.1505D}. We chose the onsite Coulomb interaction parameter $U = 4.70$ eV and onsite exchange interaction $J = 0.70$ eV, so that effective parameter $U_{\text{eff}} = 4.00$ eV, as suggested in Ref.\cite{Zhang:2013co}. Convergence test about energy cut-off, number of k-points mesh, cell size were performed\cite{sm}. In order to explore the short range and long range magnetic coupling, we substituted two In atoms with two V atoms with increasing distance. The V doped CuInTe$_2$ was simulated in $3\times 3\times 1$ supercells based on conventional cell with 144 atoms. The doped compound is CuIn$_{1-x}$V$_x$Te$_2$ with $x$ equals 5.6\%. Convergence test was also done in $2\times2\times1$ supercell, see supplementary information for details\cite{sm}. Stability of V doped CuInTe$_2$ is characterized by its formation energy, which is defined as
\begin{equation}
	E_{f} = E(\text{doped}) - n_{\text{V}} \mu_{\text{V}} - E(\text{undoped}) + n_{\text{In}} \mu_{\text{In}}.
\end{equation}
where $E$ are total energies of CuIn$_{1-x}$V$_x$Te$_2$ and CuInTe$_2$, $\mu$ are chemical potentials of V and In, $n$ is the number of corresponding atoms. In our study, different configurations have the same number of In and V atoms, so difference of formation energies between different configurations cancels the contribution of chemical potentials. And we define this difference between each configuration and that of the most stable one as relative formation energy, which equals the difference of total energies between them.

\begin{figure}[ht]
	\includegraphics[width=\columnwidth]{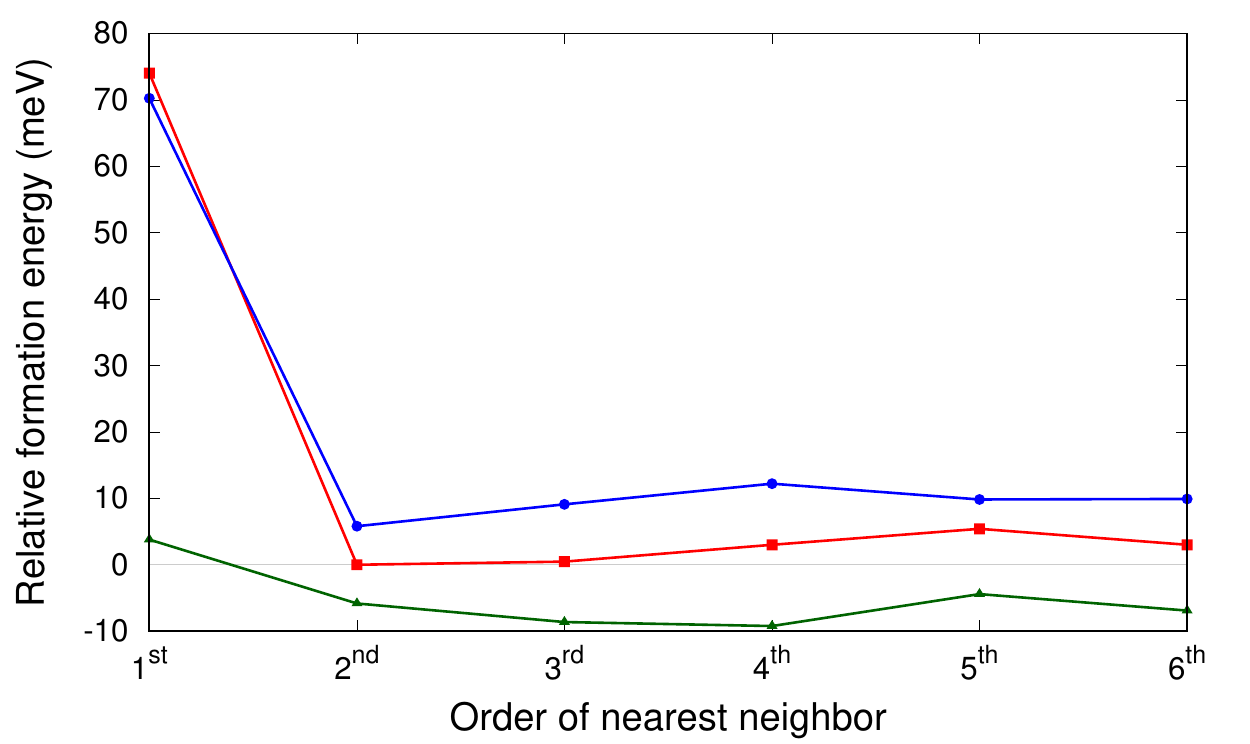}
	\caption{\label{fig:energy}The relative formation energy for V doped CuInTe$_2$in $3\times3\times1$ supercells. Red(blue) points are the relative formation energies of AFM(FM) configurations, and green ones are their difference.}
\end{figure}

\section{\label{results}Results}
The relative formation energies of V doped CuInTe$_2$ as a function of different neighboring configurations were calculated from fully relaxed $3\times3\times1$ supercell, shown in Fig.\ref{fig:energy}. The results in $2\times2\times1$ supercell and spin-orbit coupling(SOC) effect were given in supplementary material\cite{sm}. As shown in the FIG.\ref{fig:energy}, the second NN AFM state is the global minimum, with an AFM-FM energy difference 5.81 meV. The relative formation energy difference between the first NN and second NN are similar in both $2\times2\times1$ and $3\times3\times1$ supercells. Hence, subsequent calculation results of density of states and spin texture were illustrated in $2\times2\times1$ supercell.

\begin{figure}[ht]
	\includegraphics[width=\columnwidth]{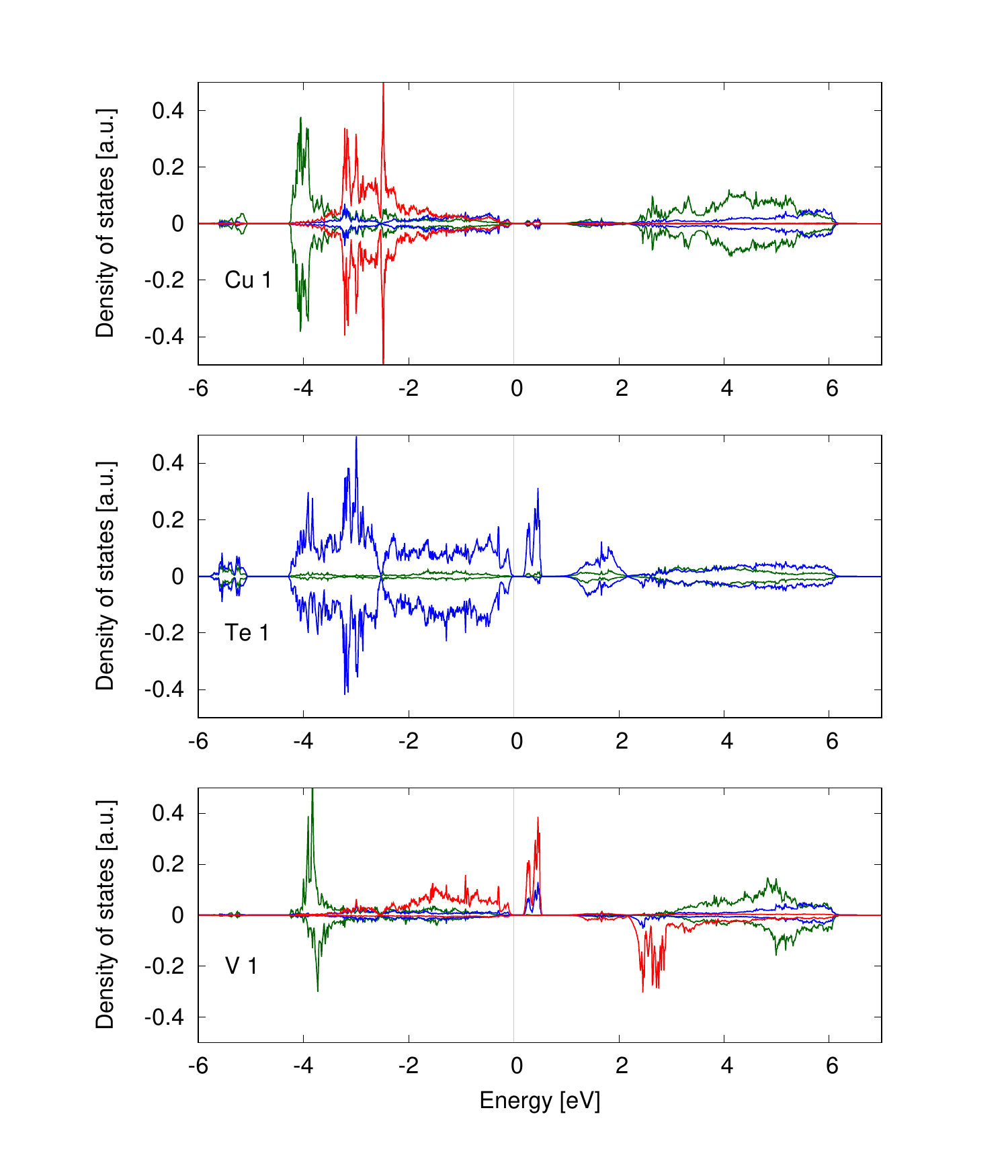}
	\caption{\label{fig:dos}The density of states of the first three atoms along the chain. Green, blue and red represent $s$, $p$ and $d$ states.}
\end{figure}

To understand the mechanism for such kind of long range carrier free AFM interaction, we calculated the projected density of states(pDOS) for second NN configuration, as shown in FIG.\ref{fig:dos} The V shows clearly polarized $p-d$ hybridized states right above the Fermi level. And the polarized $p$ states of V atoms mainly come from the hybridization of neighboring Te atoms. These polarized states have similar energy and shape as the polarized $p$ states of Te atoms. The Cu atoms, however, show no spin polarization. This is because Cu atom is at the middle of the chain, and two halves of the chain polarized it with same magnitude but opposite direction.

\begin{figure}[ht]
	\includegraphics[width=0.9\columnwidth]{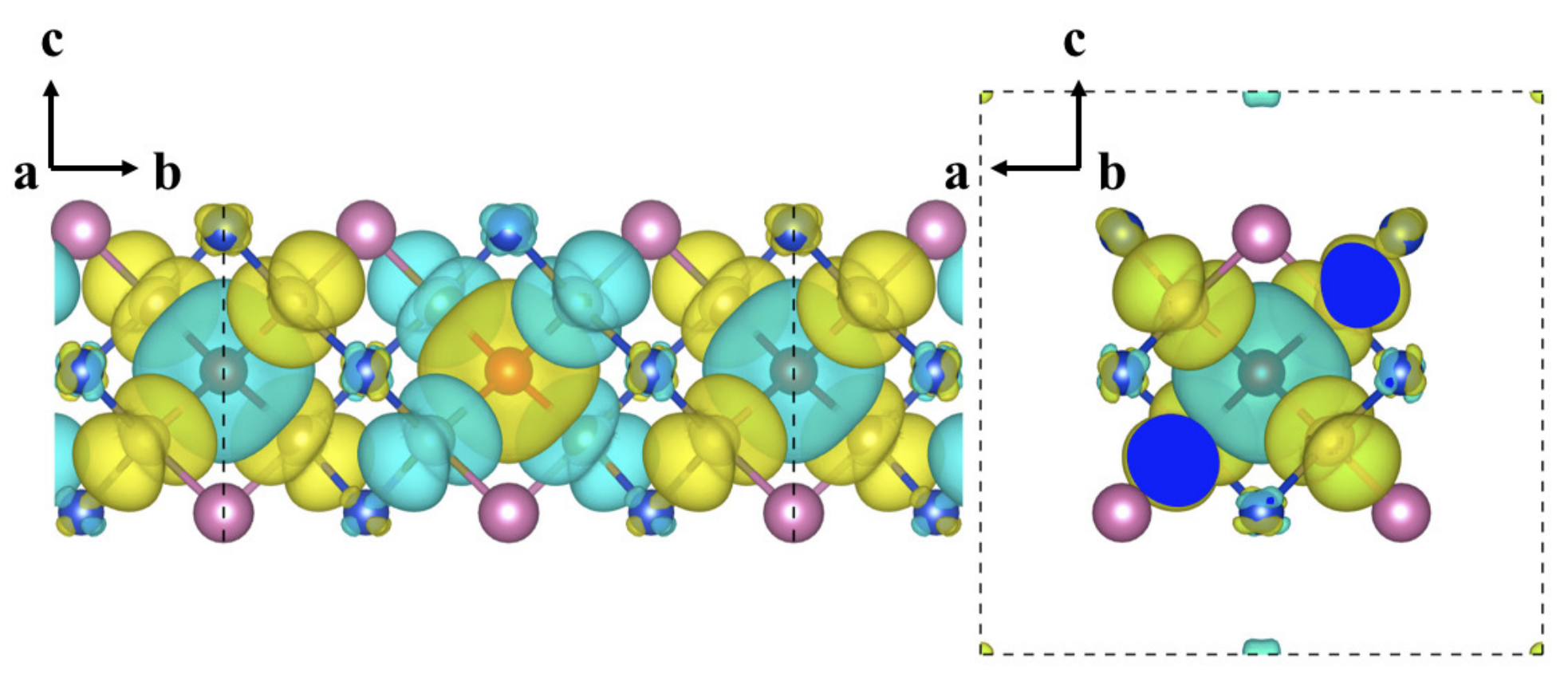}
	\caption{\label{fig:spintexture}Front view and side view of the spin texture of second NN V doped CuInTe$_2$. Dashed lines show the supercell. For clarity, atoms that are not on the chain have been removed.}
\end{figure}

In order to see the spin polarization more clearly, we calculated the polarized spin density of AFM second NN, as shown in FIG.\ref{fig:spintexture}. The V atom at the body center and another two at the face center are connected by three V-Te-Cu-Te-V chains and one V-Te-In-Te-V chain. We can see clearly polarized Cu $d$ orbitals in the middle, which serve as "stepping stones" to pass on the magnetism along the chain. In addition, long range AFM coupling was discovered in V doped AgGaTe$_2$ and AgInTe$_2$, see supplementary information for details.
	
\section{\label{model}Model}

In this section, a multi-bands Hubbard model is proposed to study the mechanism of long range AFM coupling in an A-B-C-B-A like chain which possesses the same geometry as that of the V-Te-Cu-Te-V chain, which is the building block of this AFM structure. For simplicity of theoretical modeling, we will ignore the effect of SOC in the analysis below. Nevertheless, SOC effect can still be treated using perturbation theory to achieve more accurate results, which is out the scope of this paper. We constructed various spin configurations with parallel or anti-parallel spins on $e_{g}$ states at the A(V) sites. Effective Hamiltonians in low energy subspace were derived for these two cases, and configurations with anti-parallel spins show relatively lower energy than that with parallel spins. And this accounts for an effective AFM coupling between V atoms. Finally, an estimation of the effective magnetic coupling was given.

We use the expression of onsite electron electron interaction for p and d orbitals in cubic crystal field following the result from \cite{2016PhRvB..93g5101C}. We neglected the quadrupole moment terms and derived \cite{sm}:
\begin{equation}
	\hat{V} = \hat{V}_0 + \hat{V}_{sf} + \hat{V}_{ph}.
\end{equation}
where $\hat{V}_0$, spin flipping term $\hat{V}_{sf}$ and pair hopping term $\hat{V}_{ph}$ are given by:
\begin{equation}\label{eqn:interaction}
\begin{split}
	\hat{V}_0 &= u \hat{n}^2 - v \hat{m}_z^2 - (u - v) \hat{n} + 8v \sum _{\alpha} \hat{n}_{\alpha \uparrow} \hat{n}_{\alpha \downarrow} ,\\
	\hat{V}_{sf} &= -2 v \sum _{\alpha \neq \beta, \sigma} \hat{c}_{\alpha \sigma}^{\dagger} \hat{c}_{\alpha, -\sigma} \hat{c}_{\beta, -\sigma}^{\dagger} \hat{c}_{\beta \sigma} ,\\
	\hat{V}_{ph} &= 2 v \sum _{\alpha \neq \beta} (\hat{n}_{\alpha \beta})^2 \\
	&= -2 v \sum _{\alpha \neq \beta, \sigma} \hat{c}_{\alpha \sigma}^{\dagger} \hat{c}_{\alpha, -\sigma}^{\dagger} \hat{c}_{\beta \sigma} \hat{c}_{\beta, -\sigma}.
\end{split}	
\end{equation}
where $\alpha, \beta$ are indices of local orbitals, $\sigma, \sigma'$ are indices of spin. $\hat{n}$ is the electron number operator, $\hat{m_z}$ is the $z$ component of the magnetic moment operator. $\hat{n}_{\alpha \beta} = \sum_{\sigma} \hat{c}_{\alpha \sigma}^{\dagger} \hat{c}_{\beta \sigma}$ is the onsite hopping between two orbitals. Here site label is omitted for simplicity. The parameters $u, v$ in equation \eqref{eqn:interaction} are defined as:
\begin{equation}
\begin{split}
	u &= \frac{1}{2}U - \frac{1}{4}J + \frac{5}{2}\Delta J ,\\
	v &= \frac{1}{4}J - \frac{3}{2}\Delta J.
\end{split}
\end{equation}
For $d$ orbitals, $U$ is the Coulomb interaction between $t_{2g}$ orbitals, $J$ is the average exchange splitting of $e_g$ and $t_{2g}$ orbitals and $\Delta J$ is the difference of exchange splitting between $e_g$ and $t_{2g}$ orbitals following the definitions in \cite{2016PhRvB..93g5101C}. While for $p$ orbitals, $U$ and $J$ are the Coulomb interaction and exchange splitting of three $p$ orbitals, and $\Delta J = 0$. 

Then the Hamiltonian of multi-orbital Hubbard model under tight binding approximation is:
\begin{equation}\label{eqn:Hamiltonian}
\begin{split}
	\hat{H} &= \sum _{\langle i, j \rangle} \sum _{\alpha \beta \sigma} t_{\alpha \beta}^{ij} \hat{c}_{i \alpha \sigma}^{\dagger} \hat{c}_{j \beta \sigma} + h.c. \\
	&+ \sum _{i \alpha \sigma} \epsilon _{i \alpha} \hat{c}_{i \alpha \sigma}^{\dagger} \hat{c}_{i \alpha \sigma} + \sum _{i} \hat{V}_0^i
\end{split}
\end{equation}
where $i, j = A, B, C$ are atomic sites and $t_{\alpha \beta}^{ij}$ are hopping integrals between nearest neighbor atoms. $\sum_{i} \hat{V}^i \approx \sum_{i} \hat{V}_0^i$ is the interaction according to equation \eqref{eqn:interaction} on each site, and the spin flip and pair hopping have been ignored by approximation. For detailed discussions, please see the latter part of this letter.

A and C atoms have local T$_{d}$ symmetry, so local $d$ orbital splits to 2 fold degenerated $e_{g}$ states and 3 fold degenerated $t_{2g}$ states. Te $p$ orbitals interact strongly with nearby $t_{2g}$(A, C) states which have the same symmetry and form $\sigma$ bonds. Te $p$ orbitals form only weaker $\pi$ bonds with nearby $e_{g}$ (A, C) states, which play less important role near the Fermi level \cite{1984PhRvB..29.1882J, 1984PhRvB..30.5904Y}, and we ignored these terms in this model. The signs of the hopping integrals depend on the geometry of the chain. Using the Slater-Koster matrix elements \cite{1954PhRv...94.1498S}, the signs of hopping integrals can be determined\cite{sm}. The magnitudes of hopping were approximated to be the same between A, B and B, C with $t = V_{pd\pi}/\sqrt{3}$.

Starting from the neutral state, different configurations were obtained by nearest neighbor hopping among $p$ orbitals of B and $t_{2g}$ (A, C). The $e_g$ (A, C) were fixed in this model. The relative position of each atomic level were taken to be the same as V(A), Te(B), Cu(C), and this relation can as well be applied to compounds with elements one row lower or higher. The total energies of this chain with parallel and anti-parallel initial V spins were calculated using equation \eqref{eqn:Hamiltonian} \cite{sm}.

\begin{figure}
	\includegraphics[width=\columnwidth]{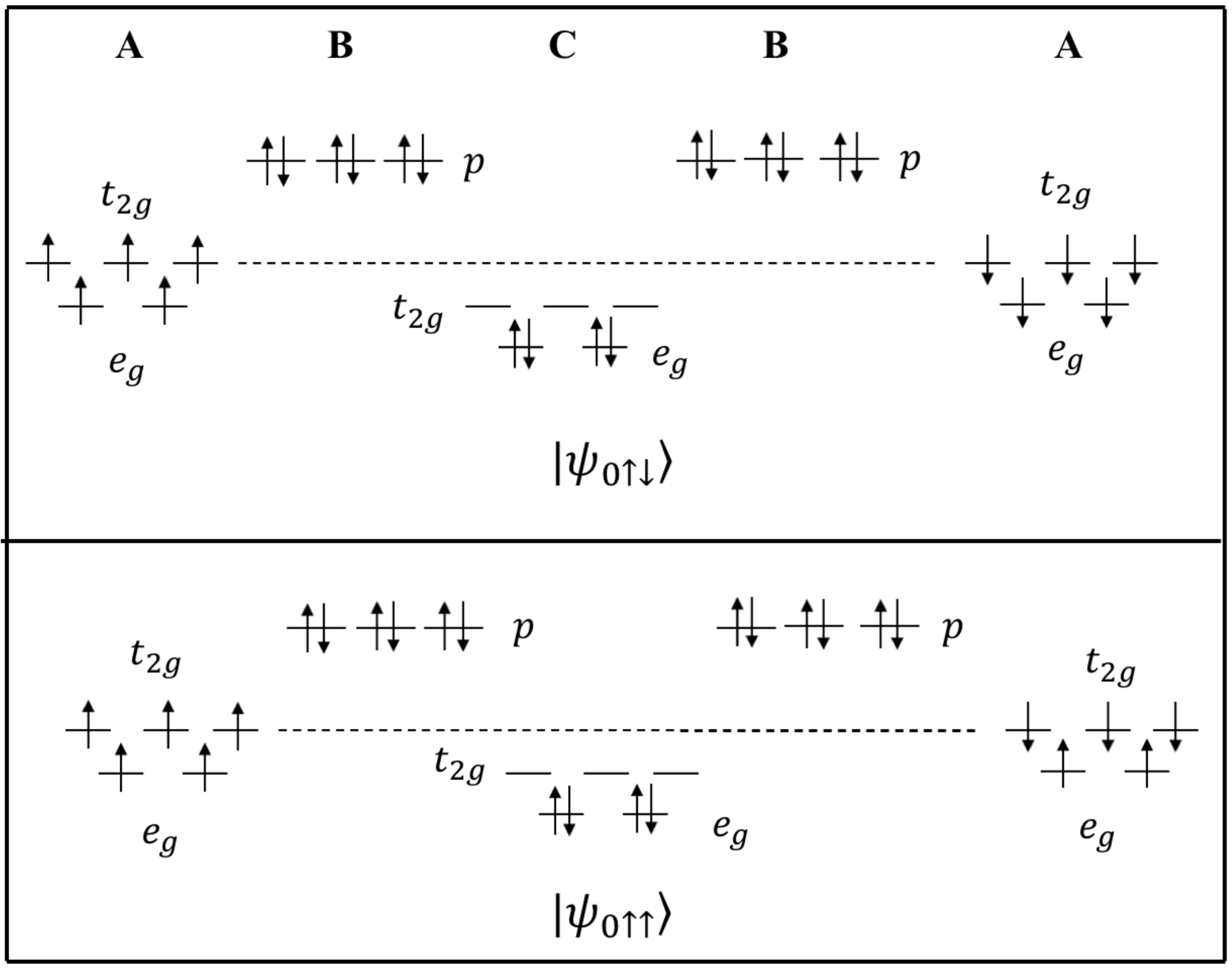}
	\caption{\label{fig:lowest}The lowest energy configurations with parallel and anti-parallel spins on $e_{g}$ states of site A(V).}
\end{figure}

The lowest energy states with parallel($\psi_{0 \uparrow \uparrow}$) and anti-parallel($\psi_{0 \uparrow \downarrow}$) $e_{g}$ spins at site A are shown in FIG.\ref{fig:lowest}. In state $\psi_{0 \uparrow \downarrow}$, two V(A) atoms have half-filled $3d$ orbitals with anti-parallel spins, two Te(B) atoms have fully filled $5p$ orbitals and Cu(C) atom has empty $t_{2g}$ orbitals and fully filled $e_g$ orbitals. The spin configuration of two A atoms in this state is $3d_{\uparrow}^5$, $3d_{\downarrow}^5$($S = \frac{5}{2}$) obeying the Hund's rule. While in state $\psi_{0 \uparrow \uparrow}$, electron filling is the same at site B, C, but one V(A) atom has $3d_{\uparrow}^5$ filling and another has $e_{g\uparrow}^2 t_{2g\downarrow}^3$ filling.  The energy of $\psi_{0 \uparrow \uparrow}$ is $24 v_A$ larger than that of $\psi_{0 \uparrow \downarrow}$. This is because $\psi_{0 \uparrow \downarrow}$ state has larger local magnetic moment and results in a smaller energy. Although hopping directly between A and C is not allowed in this model, it can happen by virtual hopping. In this process, one spin on the $p$ state at site B hops to $t_{2g}$ state at site C first, then one spin on the $t_{2g}$ state at site A hops to B site, and vise versa. This is a typical process that happens in super-exchange mechanism\cite{1950PhRv...79..350A}. We denote the intermediate states and states after one virtual hopping as $\psi_{1 \uparrow \downarrow}, \psi_{1 \uparrow \uparrow}$ and $\psi_{2 \uparrow \downarrow}, \psi_{2 \uparrow \uparrow}$. In the parallel case, two A atoms are not symmetric, so we denote state after one virtual hopping of an up(down) spin from $t_{2g}$ orbitals at site A as $\psi_{2 \uparrow \uparrow}(\psi_{2 \uparrow \uparrow}')$.

Relatively higher energy states are those with smaller local magnetic moment at site A, the $t_{2g}$ spins at site A can be different. There are huge energy barriers between these states and the two lowest energy states, and we focus only on the lowest energy states and their intermediate states, so these states are not considered in the analysis below. Starting from lowest energy configurations, higher order configurations can be obtained by nearest neighbor hopping. We can apply perturbation theory to calculate the effective ground state energy. Applying the fourth order perturbation theory, we obtain:
\begin{equation}
\begin{split}
	E_{\uparrow \downarrow} = E_{0\uparrow \downarrow} - \frac{18 t^2}{E_{1\uparrow \downarrow}-E_{0\uparrow \downarrow}} - \frac{18 t^4}{(E_{1\uparrow \downarrow}-E_{0\uparrow \downarrow})^2 (E_{2\uparrow \downarrow}-E_{0\uparrow \downarrow})}
\end{split}
\end{equation}
and
\begin{equation}
\begin{split}
	E_{\uparrow \uparrow} &= E_{0\uparrow \uparrow} - \frac{18 t^2}{E_{1\uparrow \uparrow}-E_{0\uparrow \uparrow}} \\
	&- \frac{9 t^4}{(E_{1\uparrow \uparrow}-E_{0\uparrow \uparrow})^2}(\frac{1}{E_{2\uparrow \uparrow} - E_{0\uparrow \uparrow}} + \frac{1}{E_{2\uparrow \uparrow}' - E_{0\uparrow \uparrow}}).	
\end{split}
\end{equation}
where all $E$ terms refer to the energies of states defined by the subscripts. The coefficients of the fourth order terms come from the degeneracy of $d$ orbitals of C atom. The fourth order terms are different for parallel and anti-parallel cases because two A atoms are inequivalent for the parallel case. This gives us the energy difference between parallel and anti-parallel initial V spins with forth order correction is:
\begin{equation}\label{eqn:difference}
	E_{\uparrow \uparrow} - E_{\uparrow \downarrow} = 24 v_A - \frac{72 t^4 v_A}{(\Delta E_{10})^2(\Delta E_{20})^2}	
\end{equation}
where $\Delta E_{10} = E_{1\uparrow \downarrow} - E_{0\uparrow \downarrow} = E_{1\uparrow \uparrow} - E_{0\uparrow \uparrow}$, $\Delta E_{20} = E_{2\uparrow \downarrow} - E_{0\uparrow \downarrow} = E_{2\uparrow \uparrow} - E_{0\uparrow \uparrow}$. 

We found that anti-parallel initial spins would have lower energy than that of the parallel one. So the system prefers long range AFM interaction. This long range AFM interaction is mediated by low-lying $d$ orbitals of Cu atoms. If we replace the Cu atoms on the chain by Ag atoms, the low energy configurations does not change much, while $\Delta E_{10}$ and $\Delta E_{20}$ become larger due to higher $d$ level of Ag atoms. The energy difference becomes larger according to \eqref{eqn:difference}. Remember that two V atoms are connected by six chains, three to the left, three to the right. To further illustrate the stepping stone mechanism mediated by $d$ states, we replaced the six Cu atoms by Na atoms at the stepping stone sites and found that the energy difference between FM and AFM  configurations is dramatically reduced to 1.6 meV. This small residue energy is probably due to the magnetism mediated by In atoms. We also replaced the six Cu atoms by Ag atoms at stepping stone sites and found that the energy difference between FM and AFM configurations changes from 7.57 meV for the Cu case to 8.89 meV for the Ag case, which confirms the validity of our model.

The actual energy difference is small due to the following reasons: 1) The two V atoms in this system are actually connected by three V-Te-Cu-Te-V chains. Three Cu atoms in the middle provide more hopping channels, which will enlarge the coefficients of the second term in equation \eqref{eqn:difference}; 2) configurations with different local magnetic moments also have low energies. For example, energy of configurations with $3d_{\uparrow}^4 3d_{\downarrow}^1$, $3d_{\downarrow}^4 3d_{\uparrow}^1$ (spin distribution of V atoms is $S = \frac{3}{2}$) is 32$v_A$ larger than that of lowest energy state. Although these configurations are ignored due to huge energy barriers, see supplementary information for details, they also help to reduce the magnetic coupling. And the ground state will be the superposition of these states, so that the actual magnetic moments of V atoms are smaller than that of the lowest energy configuration, which has the largest magnetic moments. Also, the effect of spin flipping and pair hopping become more important when $S < \frac{5}{2}$, because in this case there are both up spins and down spins at site A. Spin flipping happens when $d$ orbitals of A are half-filled and pair hopping happens when there are both doubly filled and empty orbitals. And spin flipping effect of "stepping stones" induce a ferromagnetic superexchange when there is a 90 degree twisted chain structure\cite{1955PhRv..100..564G}.

Our model gives clues to finding stable long range AFM order in DMS, and several guidelines can be formed: 1) To search for a long range AFM order independent of RKKY interaction, the host cell should be carrier free; 2) There should be low lying states along the path between magnetic dopants which serve as "stepping stones" to pass on the spin exchange process. Such states are relatively easy to be magnetized due to a large local Coulomb repulsion that may induce charge redistribution along the chain. As a result, long range magnetic order emerges with the assistance of the charge redistribution; 3) Also host cell should consist of large atoms, since they may suppress neighboring superexchange and favor the long range interactions.

\section{\label{conclusion}Conclusion}
To conclude, we have discovered an intrinsic long range AFM structure in V doped CuInTe$_2$. This AFM coupling can not be explained by RKKY interaction since there are no carriers in this system. In addition, this AFM coupling can not be explained by van Vleck paramagnetism based on band inversion in topological insulators because system is topologically trivial. It can rather be explained by an extended superexchange mechanism in a A-B-C-B-A like chain structure, which we named as "stepping stone" mechanism. Energy difference between each two of the three atoms, local Coulomb interaction, local exchange splitting, local orbital degeneracy and asymmetry between two halves of the chain help to induce an effective AFM coupling between two magnetic dopants. A multi-orbitals Hubbard model was proposed and confirmed by further DFT calculations. Our model can also be applied to other long range chain structures with different types of localized electronic states locating at the stepping stone sites.

\section{Acknowledgement}
We are grateful for the financial support of Chinese University of Hong Kong (CUHK) (Grant No.4053084), University Grants Committee of Hong Kong (Grant No. 24300814), and start-up funding of CUHK.

\title{Supplemental materials - Intrinsic long range antiferromagnetic coupling in dilutely V doped CuInTe2}

\author{Weiyi Gong}
\author{Ching-Him Leung}
\author{Chuen-Keung Sin}
\author{Jingzhao Zhang}
\author{Xiaodong Zhang}
\author{Bin Xi}
\author{Junyi Zhu}%
 \email{jyzhu@phy.cuhk.edu.hk}
\affiliation{%
 Department of Physics, The Chinese University of Hong Kong, Hong Kong SAR, China
}%

\maketitle
\section{Convergence test and calculation in $\text{AgGaTe}_2$ and $\text{AgInTe}_2$}

\begin{figure}[ht]
	\includegraphics[width=\columnwidth]{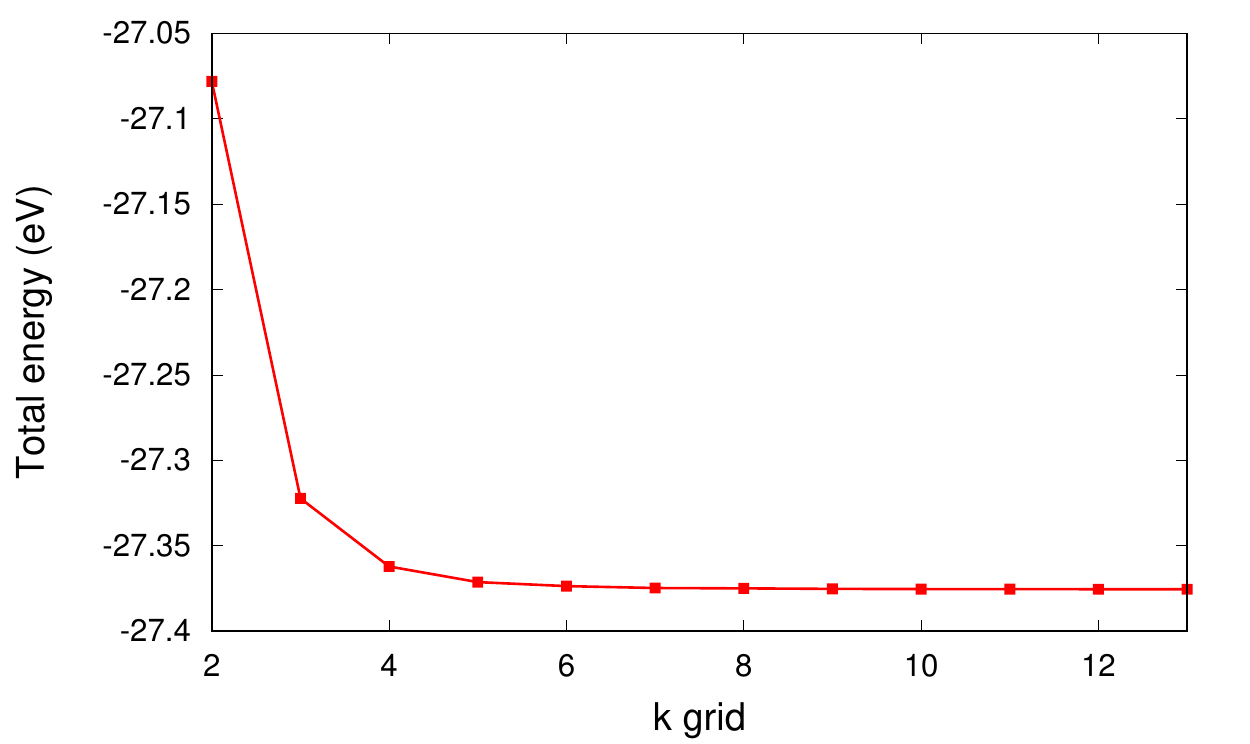}
	\caption{\label{fig:ktest}The convergence test of k grid. For each $i$ in the horizontal axis, the k grid is $i \times i \times i$.}
\end{figure}

All calculations were performed using projected augmented plane wave method\cite{s1994PhRvB..5017953B} and density functional theory with Perdew-Burke-Ernzerhof generalized gradient approximation(GGA) \cite{s2009PhRvL.102c9902P} as implemented in VASP code\cite{s1996PhRvB..5411169K}. The V doped CuInTe$_2$ was simulated in $2\times2\times1$ supercell and $3\times3\times1$ supercell. Atoms were relaxed with force tolerance of 0.01 eV\AA$^{-1}$. A plane wave energy cut-off of 300 eV was used in all calculations. The Brillouin zone integrations were performed by using $\Gamma$ centered  $5\times5\times5$ k-points grid. The convergence test of energy of the system using different number of k points was shown in figure \ref{fig:ktest}. 

\begin{figure}
	\includegraphics[width=\columnwidth]{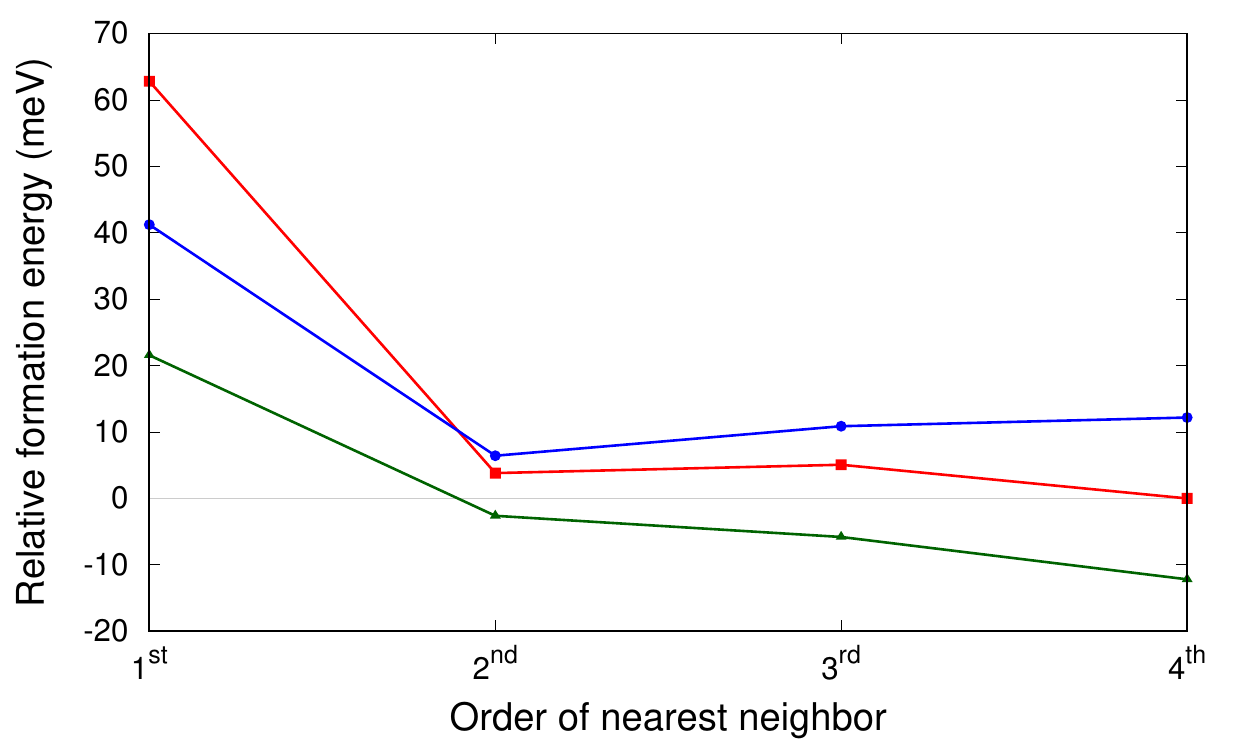}
	\includegraphics[width=\columnwidth]{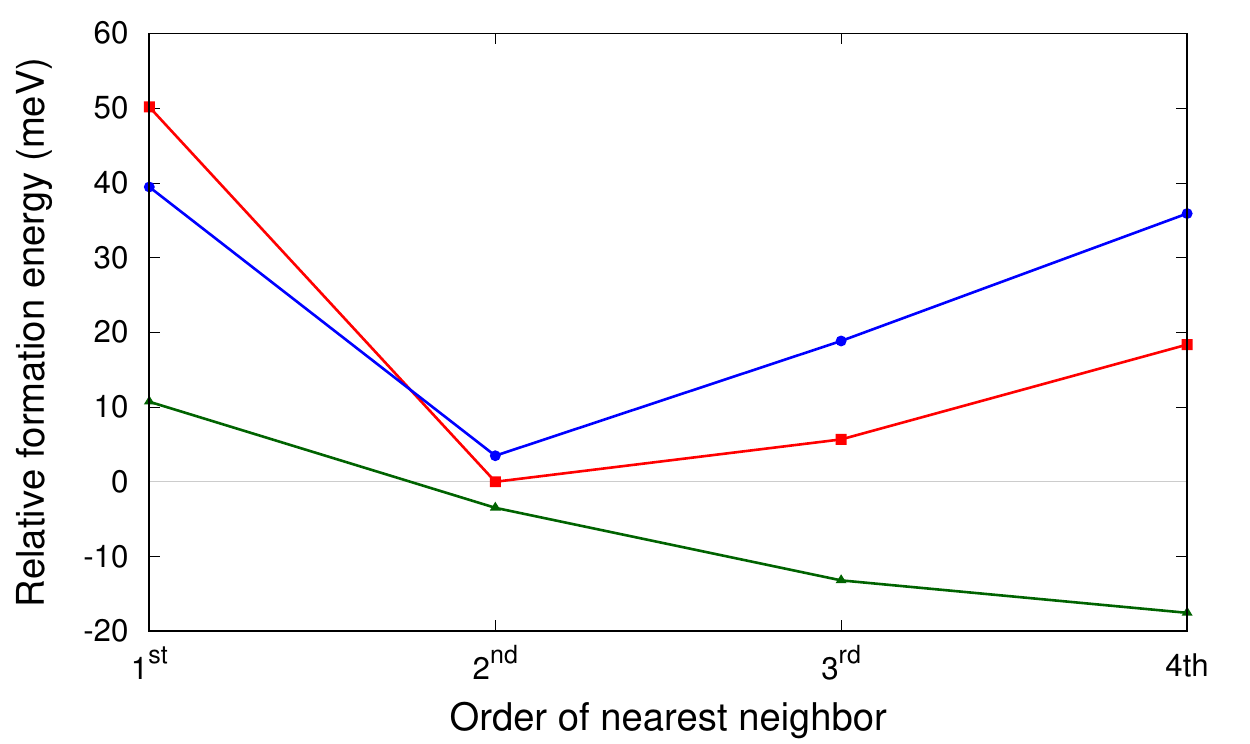}
	\caption{\label{fig:eat}The relative formation energy of different nearest neighbor configurations in V doped AgGaTe$_2$ and AgInTe$_2$. Red(blue) points are the relative formation energies of AFM(FM) configurations, and green ones are their difference.}
\end{figure}

Similar calculations of the relative formation energy of different configurations were done in AgGaTe$_2$ and AgInTe$_2$ in $2\times2\times1$ supercells, as shown in figure \ref{fig:eat}. Similar as in CuInTe$_2$, V atoms in AgInTe$_2$ have a lowest energy in the second nearest neighbor configuration, while lowest energy configuration in AgGaTe$_2$ is the forth nearest neighbor. 

\section{Calculations in $2\times2\times1$ supercell}
\begin{figure}
	\includegraphics[width=\columnwidth]{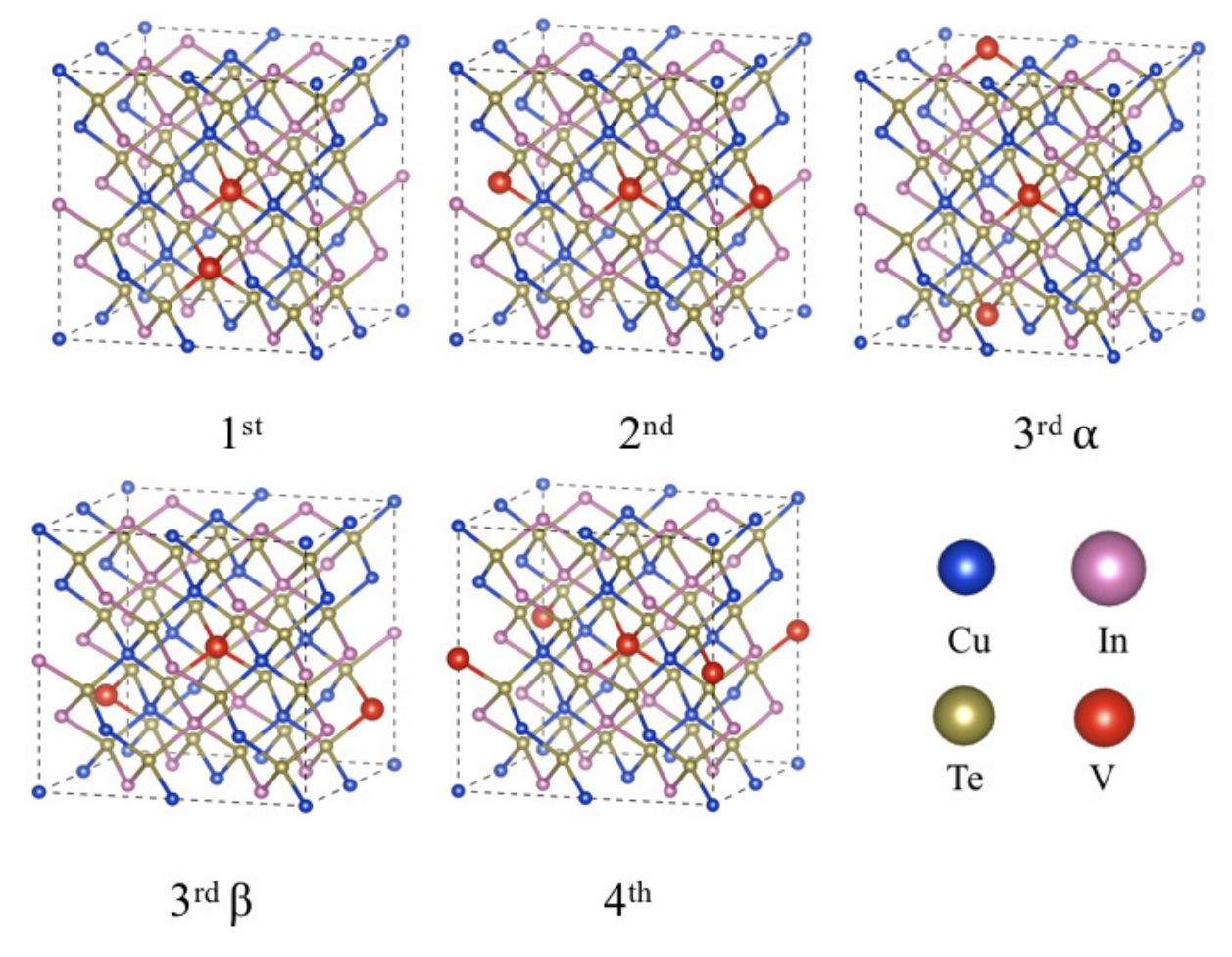}
	\caption{\label{config}The nearest neighboring configurations of two V atoms doped into the $2\times2\times1$ supercell.}
\end{figure}

\begin{figure}
	\includegraphics[width=\columnwidth]{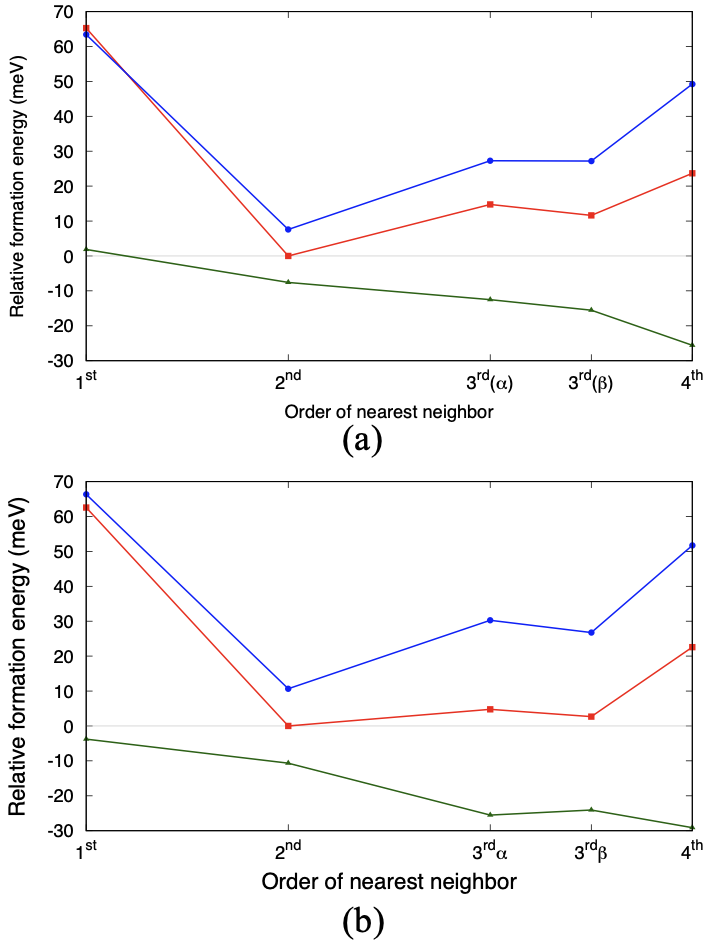}
	\caption{\label{2x2x1}The relative formation energy in $2\times2\times1$ supercell. (a) without SOC; (b) with SOC. Red(blue) line represents the formation energy of AFM(FM) configuration, green line represents their difference.}
\end{figure}

The calculation setup in $2\times2\times1$ supercell was the same as stated in the main text. An In atom at body center of the supercell was substituted with a V atom. Another V atom replaces an In atom at various neighboring sites to the center V atom, from the first nearest neighbor(NN) to the forth NN, as shown in Fig.\ref{config}. Cu, V and In atoms have approximately local $T_d$ symmetry in the host cell, which is a property of chalcopyrite. The point group of this supercell is $\bar{4}2d$\space ($D_{2d}$). Under this symmetry, the third NN has two nonequivalent configurations, while other three have only one configuration for each. 
Due the limitation of calculation resources, SOC calculation was only done in $2\times2\times1$ supercell. The relative formation energy without and with SOC is shown in FIG.\ref{2x2x1}.

Although the effect of spin orbit coupling (SOC) is ignored in the theoretical model in the main text, SOC may induce further splitting of energy levels of heavy atoms like Te. Due to the SOC effect, the $p$ level of Te will become lower, while this effect at Cu and In atoms is not so obvious. Hence, according to the perturbation theory, smaller energy difference will cause a larger magnetic coupling. The results of DFT calculation with SOC effect considered confirmed this point. At the second nearest neighbor, AFM has energy 7.57 meV lower than that of FM without SOC effect, while this value becomes 10.74 meV with SOC considered. Although the magnetic coupling of the first NN changes from FM to AFM once SOC is considered, the general trends of formation energy as functions of neighboring sites are similar.

\section{Calculations in $3\times3\times1$ supercell}
The setup is the same as above, but there are more NN configurations. In FIG.\ref{3x3x1}, In atoms are labeled by ($n$, $m$), where $n$ is the label of cell and $m$ is the label of In atom. In the magnetic doping process, two In atoms are replaced by two V atoms. Due to the symmetry in chalcopyrite, we fix one V atom at position (1, 6). The position of the second V atom is shown in TAB.\ref{tab:nn}.

\begin{table}[ht]
\caption{\label{tab:nn}The positions of the second V atom.}
\begin{ruledtabular}
\begin{tabular}{ccccccc}
	NN&1&2&3&4&5&6\\
	\hline\\
	position&(2, 8)&(2, 6)&(5, 1) or (5, 3)&(5, 6)&(3, 8)&(6, 1)\\
	\hline\\
	V-V distance(\AA) &4.38&6.19&7.58&8.75&9.79&11.59\\
\end{tabular}
\end{ruledtabular}
\end{table}

\begin{figure}
	\includegraphics[width=\columnwidth]{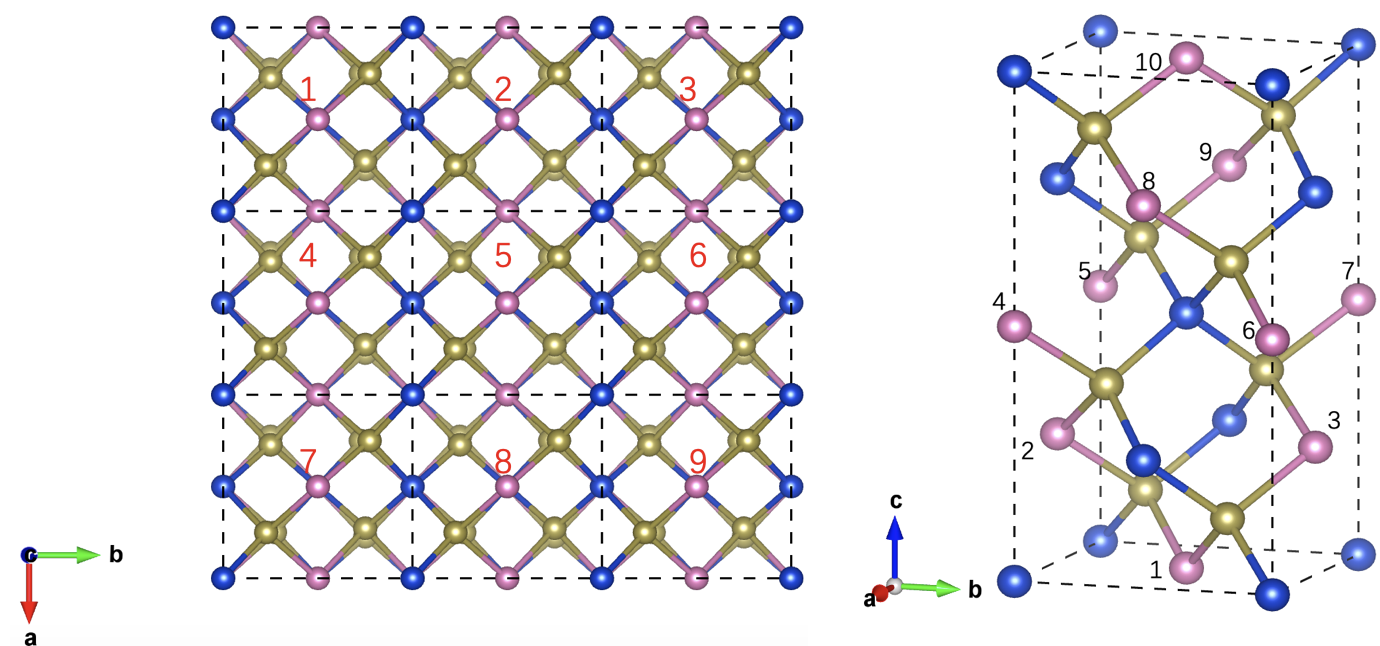}
	\caption{\label{3x3x1}The nearest neighboring configurations of two V atoms doped into the $3\times3\times1$ supercell.}
\end{figure}

The setup of calculation in $3\times 3\times 1$ supercell was the same as that of $2\times 2\times 1$ supercell. Static calculation with SOC was performed to qualitatively check the effect of SOC. The results without and with static SOC calculation were shown in FIG.\ref{3x3x1}. The general trends of the formation energy and energy difference between FM and AFM are similar. We see that the second NN is still the most stable configuration in larger supercell without SOC, while the stability of the sixth NN is comparable with that of the second NN. The formation energy is determined both by magnetic coupling strength and the local stress that may slightly vary with or without SOC. Nevertheless, the stabilization of long range configurations versus the first nearest neighboring configuration is valid for both setup.
\begin{figure}
	\includegraphics[width=\columnwidth]{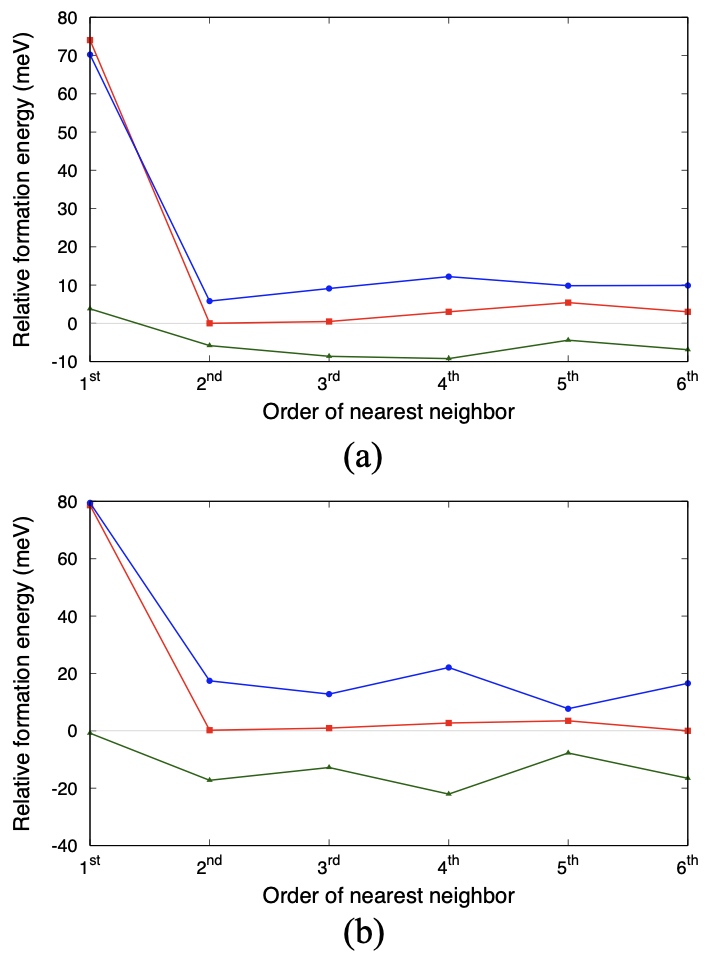}
	\caption{\label{3x3x1}The relative formation energy in $3\times 3\times1$ supercell. (a) without SOC; (b) with SOC. Red(blue) line represents the formation energy of AFM(FM) configuration, green line represents their difference.}
\end{figure}

\section{\label{Simplifications}Normal ordering of local operators}

The electron number operator, onsite hopping and local magnetic moment operator at each site are defined as:
\begin{equation}
\begin{split}
	\hat{n} &= \sum _{\alpha \sigma} \hat{c}_{\alpha \sigma}^{\dagger} \hat{c}_{\alpha \sigma} = \sum _{\sigma} \hat{n}_{\alpha} ,\\
	\hat{n}_{\alpha \beta} &= \sum _{\sigma} \hat{c}_{\alpha \sigma}^{\dagger} \hat{c}_{\beta \sigma} ,\\
	\hat{\bm{m}} &= \sum _{\alpha \sigma \sigma'} \hat{c}_{\alpha \sigma}^{\dagger}\bm{\tau}_{\sigma \sigma'} \hat{c}_{\alpha \sigma'} ,\\
	\bm{\tau}_{\sigma \sigma'} &= (\tau _{\sigma \sigma'}^x, \tau _{\sigma \sigma'}^y, \tau _{\sigma \sigma'}^z).
\end{split}
\end{equation}
where $\alpha, \beta$ are indices of local orbitals, $\sigma, \sigma'$ are indices for spins and $\tau ^x, \tau ^y, \tau ^z$ are three Pauli matrices. The normal ordering of $\hat{n}^2$ and $\hat{n}_{\alpha \beta}^2$ are defined as:
\begin{equation}\label{eqn:n2normal_simplified}
\begin{split}
	:\mathrel{\hat{n}^2}: &= \sum _{\alpha \beta \sigma \sigma'} \hat{c}_{\alpha \sigma}^{\dagger} \hat{c}_{\beta \sigma'}^{\dagger} \hat{c}_{\beta \sigma'} \hat{c}_{\alpha \sigma} \\
	&= -\sum _{\alpha \beta \sigma \sigma'} \hat{c}_{\alpha \sigma}^{\dagger} (\delta _{\alpha \beta} \delta _{\sigma \sigma'} - \hat{c}_{\alpha \sigma} \hat{c}_{\beta \sigma'}^{\dagger}) \hat{c}_{\beta \sigma'} \\
	&= -\hat{n} + \hat{n}^2,
\end{split}
\end{equation}
and 
\begin{equation}\label{eqn:nab}
\begin{split}
	:\mathrel{\hat{n}_{\alpha \beta}^2}: &= \sum _{\sigma \sigma'} \hat{c}_{\alpha \sigma}^{\dagger} \hat{c}_{\alpha \sigma'}^{\dagger} \hat{c}_{\beta \sigma'} \hat{c}_{\beta \sigma} \\
	&= -\sum _{\sigma \sigma'} \hat{c}_{\alpha \sigma}^{\dagger} (\delta_{\alpha \beta} \delta_{\sigma \sigma'} - \hat{c}_{\beta \sigma} \hat{c}_{\alpha \sigma'}^{\dagger}) \hat{c}_{\beta \sigma'} \\
	&= -\hat{n}_{\alpha \beta} \delta_{\alpha \beta} + \hat{n}_{\alpha \beta}^2.
\end{split}
\end{equation}
Note that  $[\hat{n}_{\alpha \sigma}, \hat{n}_{\alpha \sigma'}] = 0$ and $\hat{n}_{\alpha \sigma}^2 = \hat{n}_{\alpha \sigma}$, so we have:
\begin{equation}\label{eqn:nab1}
\begin{split}
	\hat{n}_{\alpha}^2 &= (\hat{n}_{\alpha \uparrow} + \hat{n}_{\alpha \downarrow})^2 \\
	&= \hat{n}_{\alpha \uparrow}^2 + \hat{n}_{\alpha \downarrow}^2 + \hat{n}_{\alpha \uparrow} \hat{n}_{\alpha \downarrow} + \hat{n}_{\alpha \downarrow} \hat{n}_{\alpha \uparrow} \\
	&= \hat{n}_{\alpha \uparrow} + \hat{n}_{\alpha \downarrow} + 2 \hat{n}_{\alpha \uparrow} \hat{n}_{\alpha \downarrow}.
\end{split}
\end{equation}
Combining \eqref{eqn:nab} and \eqref{eqn:nab1} gives us:
\begin{equation}\label{eqn:nab2normal_simplified}
\begin{split}
	\sum _{\alpha \beta} :\mathrel{\hat{n}_{\alpha \beta}^2}: &= -\sum _{\alpha} \hat{n}_{\alpha} + \sum _{\alpha \beta} \hat{n}_{\alpha \beta}^2 \\
	&= - \hat{n} + \sum _{\alpha} \hat{n}_{\alpha}^2 + \sum _{\alpha \neq \beta} \hat{n}_{\alpha \beta}^2 \\
	&= 2 \sum _{\alpha} \hat{n}_{\alpha \uparrow} \hat{n}_{\alpha \downarrow} + \sum _{\alpha \neq \beta} \hat{n}_{\alpha \beta}^2.
\end{split}
\end{equation}

Using the relation $\bm{\tau}_{\sigma \sigma'} \cdot \bm{\tau}_{\zeta \zeta'} = 2 \delta _{\sigma \zeta'} \delta _{\sigma' \zeta} - \delta _{\sigma \sigma'} \delta _{\zeta \zeta'}$, we have:
\begin{equation}\label{eqn:m}
\begin{split}
	\hat{\bm{m}}^2 &= \sum _{\alpha \sigma \sigma'} \sum _{\beta \zeta \zeta'} \bm{\tau}_{\sigma \sigma'} \cdot \bm{\tau}_{\zeta \zeta'} \hat{c}_{\alpha \sigma}^{\dagger} \hat{c}_{\alpha \sigma'} \hat{c}_{\beta \zeta}^{\dagger} \hat{c}_{\beta \zeta'} \\
	&= 2 \sum _{\alpha \beta \sigma' \sigma} \hat{c}_{\alpha \sigma}^{\dagger} \hat{c}_{\alpha \sigma'} \hat{c}_{\beta \sigma}^{\dagger} \hat{c}_{\beta \sigma'} - \hat{n}^2.
\end{split}
\end{equation}

Note that the $z$ component of magnetic moment operator satisfies:
\begin{equation}\label{eqn:mz}
\begin{split}
	\hat{m}_{z} &= \sum _{\alpha} (\hat{n}_{\alpha \uparrow} - \hat{n}_{\alpha \downarrow}) ,\\
	\hat{n}^2 + \hat{m}_{z}^2 &= 2 \sum _{\alpha \beta} (\hat{n}_{\alpha \uparrow} \hat{n}_{\beta \uparrow} + \hat{n}_{\alpha \downarrow} \hat{n}_{\beta \downarrow}).
\end{split}
\end{equation}

Combining \eqref{eqn:m} and \eqref{eqn:mz} gives us:
\begin{equation}
	\hat{m}^2 = \hat{m}_z^2 + 2 \sum _{\alpha \beta \sigma} \hat{c}_{\alpha \sigma}^{\dagger} \hat{c}_{\alpha, -\sigma} \hat{c}_{\beta, -\sigma}^{\dagger} \hat{c}_{\beta \sigma}.
\end{equation}

The $\alpha = \beta$ terms in the above sum is :
\begin{equation}
\begin{split}
	&2 \sum _{\alpha} (\hat{c}_{\alpha \uparrow}^{\dagger} \hat{c}_{\alpha \downarrow} \hat{c}_{\alpha \downarrow}^{\dagger} \hat{c}_{\alpha \uparrow} + \hat{c}_{\alpha \downarrow}^{\dagger} \hat{c}_{\alpha \uparrow} \hat{c}_{\alpha \uparrow}^{\dagger} \hat{c}_{\alpha \downarrow}) \\
	&= 2 \sum _{\alpha} [\hat{n}_{\alpha \uparrow}(1-\hat{n}_{\alpha \downarrow}) + \hat{n}_{\alpha \downarrow}(1-\hat{n}_{\alpha \uparrow})] \\
	&= 2 \hat{n} - 4 \sum _{\alpha} \hat{n}_{\alpha \uparrow} \hat{n}_{\alpha \downarrow},
\end{split}
\end{equation}
so 
\begin{equation}\label{eqn:m2_simplified}
	\hat{\bm{m}}^2 = \hat{m}_z^2 + 2 \sum _{\alpha \neq \beta, \sigma} \hat{c}_{\alpha \sigma}^{\dagger} \hat{c}_{\alpha, -\sigma} \hat{c}_{\beta, -\sigma}^{\dagger} \hat{c}_{\beta \sigma} + 2\hat{n} - \sum _{\alpha} \hat{n}_{\alpha \uparrow} \hat{n}_{\alpha \downarrow}.
\end{equation}

And the normal ordering of $\hat{\bm{m}}^2$ is:
\begin{equation}\label{eqn:m2normal_simplified}
\begin{split}
		:\mathrel{\hat{\bm{m}}^2}: &= 2\sum _{\alpha \beta \sigma \sigma'} \hat{c}_{\alpha \sigma}^{\dagger} \hat{c}_{\beta \sigma'}^{\dagger} \hat{c}_{\beta \sigma} \hat{c}_{\alpha \sigma'} - :\mathrel{\hat{n}^2}: \\
		&= -2\sum _{\alpha \beta \sigma \sigma'} \hat{c}_{\alpha \sigma} (\delta_{\alpha \beta} - \hat{c}_{\alpha \sigma'} \hat{c}_{\beta \sigma'}^{\dagger}) \hat{c}_{\beta \sigma} - \hat{n}^2 + \hat{n} \\
		&= -2\sum _{\alpha \sigma \sigma'} \hat{c}_{\alpha \sigma}^{\dagger} \hat{c}_{\alpha \sigma} + \hat{\bm{m}}^2 + \hat{n} \\
		&= -3 \hat{n} + \hat{\bm{m}}^2.
\end{split}
\end{equation}

Finally, all the normal ordering of local operators have been expressed using local operators in equation \eqref{eqn:n2normal_simplified}, \eqref{eqn:nab2normal_simplified}, \eqref{eqn:m2_simplified} and \eqref{eqn:m2normal_simplified}.

\section{\label{Hamiltonian}Interacting Hamiltonian for p and d orbitals}

The electron-electron interaction for d orbitals can be expressed as\cite{s2016PhRvB..93g5101C}: 
\begin{equation}\label{eqn:V}
\begin{split}
	\hat{V} &= \frac{1}{2}[(U-\frac{1}{2}J+5\Delta J):\mathrel{\hat{n}^2}:-\frac{1}{2}(J-6\Delta J):\mathrel{\hat{\mathbf{m}}^2}: \\
	&+(J-6 \Delta J)\sum_{\alpha \beta}:\mathrel{(\hat{n}_{\alpha \beta})^2}:]
\end{split}
\end{equation}
where the quadrupole moment operator has been ignored since $\Delta J$ is approximately of magnitude one order lower than that of $J$\cite{sPhysRevB.29.314}. The interaction for p orbitals can be obtained be setting $\Delta J = 0$.

$U$ is the Coulomb interaction between $t_{2g}$ orbitals, $J$ is the average exchange splitting of $e_g$ and $t_{2g}$ orbitals and $\Delta J$ is the difference of exchange splitting between $e_g$ and $t_{2g}$ orbitals. Using the conventional indices (1,2,3,4,5) to represent d orbitals ($3z^2-r^2, zx, yz, xy, x^2-y^2$), respectively, the parameters can be written as:
\begin{equation}
\begin{split}
	U &= V_{23,23} ,\\
	J &= \frac{1}{2} (V_{15,51} + V_{23,32}) ,\\
	\Delta J &= V_{15,51} - V_{23,32}.
\end{split}
\end{equation}

Plug equations \eqref{eqn:n2normal_simplified}, \eqref{eqn:m2normal_simplified} and \eqref{eqn:nab2normal_simplified} into \eqref{eqn:V}, we have:
\begin{equation}\label{eqn:interaction}
	\hat{V} = \hat{V}_0 + \hat{V}_{sf} + \hat{V}_{ph} .
\end{equation}
where $\hat{V}_0$, spin flipping terms $\hat{V}_{sf}$ and pair hopping terms $\hat{V}_{ph}$ are given by:
\begin{equation}\label{eqn:interactionterms}
\begin{split}
	\hat{V}_0 &= u \hat{n}^2 - v \hat{m}_z^2 - (u - v) \hat{n} + 8v \sum _{\alpha} \hat{n}_{\alpha \uparrow} \hat{n}_{\alpha \downarrow} ,\\
	\hat{V}_{sf} &= -2 v \sum _{\alpha \neq \beta, \sigma} \hat{c}_{\alpha \sigma}^{\dagger} \hat{c}_{\alpha, -\sigma} \hat{c}_{\beta, -\sigma}^{\dagger} \hat{c}_{\beta \sigma} ,\\
	\hat{V}_{ph} &= 2 v \sum _{\alpha \neq \beta} (\hat{n}_{\alpha \beta})^2 \\
	&= -2 v \sum _{\alpha \neq \beta, \sigma} \hat{c}_{\alpha \sigma}^{\dagger} \hat{c}_{\alpha, -\sigma}^{\dagger} \hat{c}_{\beta \sigma} \hat{c}_{\beta, -\sigma}.
\end{split}	
\end{equation}
 And where site label is omitted for simplicity. The parameters $u, v$ in equation \eqref{eqn:interactionterms} are defined as:
\begin{equation}
\begin{split}
	u &= \frac{1}{2}U - \frac{1}{4}J + \frac{5}{2}\Delta J ,\\
	v &= \frac{1}{4}J - \frac{3}{2}\Delta J.
\end{split}
\end{equation}

Now the Hamiltonian of multi-orbital Hubbard model under tight binding approximation is:
\begin{equation}
\begin{split}
	\hat{H} &= \hat{H}_{t} + \hat{H}_{e} + \hat{H}_{int} ,\\
	&= \sum _{\langle i,j \rangle} \sum _{\alpha \beta \sigma} t_{\alpha \beta}^{ij} \hat{c}_{i \alpha \sigma}^{\dagger} \hat{c}_{j \beta \sigma} + h.c. \\
	&+ \sum _{i \alpha \sigma} \epsilon _{i \alpha} \hat{c}_{i \alpha \sigma}^{\dagger} \hat{c}_{i \alpha \sigma} + \sum _{i} \hat{V}^i
\end{split}
\end{equation}
where the hopping integrals are only nonzero between nearest neighbor atomic levels and their magnitudes and signs are discussed in the next section \ref{hopping}. $\hat{V}^i$ is the interacting Hamiltonian of each site according to equation \eqref{eqn:interaction}.

\section{\label{hopping}The hopping integrals}

The $p$ and $d$ orbitals are irreducible representation of spherical symmetry with angular momentum $l=1$ and $l=2$. However, in a local environment with $T_d$ symmetry, which is the local symmetry of V atoms and Cu atoms in our case, some of atomic orbitals become reducible, and will split according to the group theory. As shown in table \ref{tab:split}, there are 5 irreducible representations of $T_d$ group, and $p(l=1)$ and $d(l=2)$ become reducible and split into linear combination of the irreducible representations. 

\begin{figure}[ht]
	\includegraphics[width=\columnwidth]{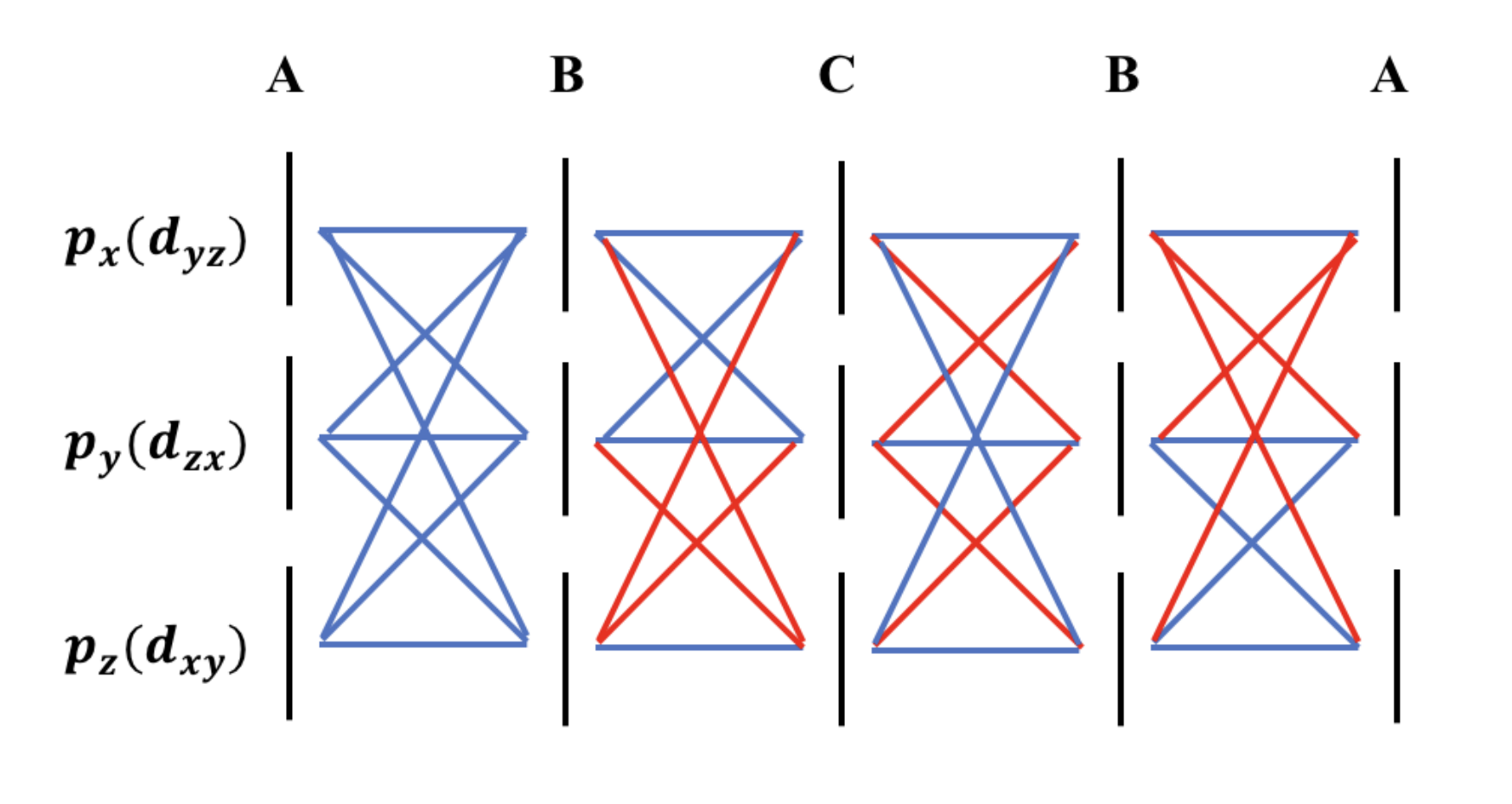}
	\caption{\label{fig:hopping}The sign of hopping integrals. Black intervals represent $p_x(d_{yx}), p_y(d_{zx}), p_z(d_{xy})$ orbitals of B(A,C) atoms. Red(Blue) lines represent that the hopping integral is positive(negative) $t$. The chain that we chose was along the direction: A-B: (1,1,1), B-C:(1,1,-1), C-B:(-1,1,-1), B-A:(-1,1,1).}
\end{figure}

\begin{table}[ht]
\caption{\label{tab:split} The character table of $T_d$ group.}
\begin{ruledtabular}
\begin{tabular}{cccccc}
	$T_d$ & $E$ & $8C_3$ & $3C_2$ & $6\sigma_d$ & $6S_4$\\
	\hline
	$A_1$ & 1 & 1 & 1 & 1 & 1\\
	$A_2$ & 1 & 1 & 1 & -1 & -1\\
	$E$ & 2 & -1 & 2 & 0 & 0\\
	$T_1$ & 3 & 0 & -1 & -1 & 1\\
	$T_2$ & 3 & 0 & -1 & 1 & -1\\
	\hline
	$\Gamma_p$ & 3 & 0 & -1 & 1 & -1\\
	$\Gamma_d$ & 5 & -1 & 1 & 1 & -1
\end{tabular}
\end{ruledtabular}
\end{table}
From the character table we get the splitting relations:
\begin{eqnarray}
	\Gamma_p &= T_2\\
	\Gamma_d &= E + T_2
\end{eqnarray}
So the $p$ orbitals in $T_d$ symmetry only interact with $t_{2g}$ states of $d$ orbitals, and interact only weakly with $e_g$ states \cite{s1984PhRvB..29.1882J}. The magnitude and signs of hopping matrix elements for each pair of $p$ orbitals and $t_{2g}$ orbitals can be determined by the Slater Koster matrix \cite{s1954PhRv...94.1498S}. The magnitude is $t = V_{pd\sigma}/\sqrt{3}$, while the signs are shown in FIG. \ref{fig:hopping}. 

\section{\label{energy}Low energy configurations}

For anti-parallel case, the lowest energy configuration has energy:
\begin{equation}
\begin{split}
	E_{0\uparrow \downarrow} &= 4 \epsilon_{A1} + 6 \epsilon_{A2} + 12 \epsilon_{B} + 4 \epsilon_{C1} \\
	&+ 40 u_A - 40 v_A + 60 u_B + 60 v_B + 12 u_C + 20 v_C.
\end{split}
\end{equation}
where $A1$, $C1$ and $A2$, $C2$ are corresponding $e_g$ and $t_{2g}$ states. In this configuration, two V(A) atoms have half-filled $3d$ orbitals, two Te(B) atoms have fully filled $5p$ orbitals and Cu(C) atom has empty $t_{2g}$ orbitals and fully filled $e_g$ orbitals. The spin configuration of two A atoms in this state is $3d_{\uparrow}^5$, $3d_{\downarrow}^5$ obeying the Hund's rule. Electrons on $t_{2g}$ orbitals of A can effectively hop to $t_{2g}$ orbitals of C mediated by $p$ electrons on B. In this process, one B electron hop to C first, and one A electron hop to B next. The energy of intermediate states and the states after one effective hopping are:
\begin{equation}
\begin{split}
	E_{1\uparrow \downarrow} &= E_{0\uparrow \downarrow} + 8 u_C - 10 u_B - 10 v_B - \Delta_{BC} + \delta_C ,\\
	E_{2\uparrow \downarrow} &= E_{0\uparrow \downarrow} + 8 u_C - 8 u_A + 8 v_A - \Delta_{AC} + \delta_C - \delta_A.	
\end{split}
\end{equation}
where $\Delta_{AC} = \epsilon_{A1} - \epsilon_{C1}$ and $\Delta_{BC} = \epsilon_{B} - \epsilon_{C1}$ are energy difference between A, C and B $p$, C $e_g$ states. And $\delta_A$ and $\delta_C$ are the crystal field splitting of A and C atoms. Applying the fourth order perturbation theory gives us:
\begin{equation}
	E_{\uparrow \downarrow} = E_{0\uparrow \downarrow} - \frac{18 t^2}{E_{1\uparrow \downarrow}-E_{0\uparrow \downarrow}} - \frac{18 t^4}{(E_{1\uparrow \downarrow}-E_{0\uparrow \downarrow})^2 (E_{2\uparrow \downarrow}-E_{0\uparrow \downarrow})}.
\end{equation}
where the coefficients in front of each terms come from the degeneracy of $p$ and $d$ orbitals. 

Similar procedure can be applied to parallel initial spins of V atoms. The lowest energy is:
\begin{equation}
\begin{split}
	E_{0\uparrow \uparrow} &= 4 \epsilon_{A1} + 6 \epsilon_{A2} + 12 \epsilon_{B} + 4 \epsilon_{C1} \\
	&+ 40 u_A - 16 v_A + 60 u_B + 60 v_B + 12 u_C + 20 v_C.
\end{split}
\end{equation}

We find that it is $24v_A$ lower than $E_{\uparrow \downarrow}$. This is because $\psi_{0 \uparrow \downarrow}$ state has larger local magnetic moment and results in a smaller energy. Electrons at sites A can hop to site C via a similar effective hopping process. In the parallel case, two A atoms are asymmetric, so we denote the state after one effective hopping of an up(down) spin from $t_{2g}$ orbitals at site A as $\psi_{2 \uparrow \uparrow}(\psi_{2 \uparrow \uparrow}')$. Starting from the lowest energy configuration, we listed the energies of the states with the second and forth order corrections:
\begin{equation}
\begin{split}
	E_{1\uparrow \uparrow} &= E_{1\uparrow \uparrow}' = E_{0\uparrow \uparrow} + 8 u_C - 10 u_B - 10 v_B - \Delta_{BC} + \delta_C ,\\
	E_{2\uparrow \uparrow} &= E_{0\uparrow \uparrow} + 8 u_C - 8 u_A + 8 v_A - \Delta_{AC} + \delta_C - \delta_A ,\\
	E_{2\uparrow \uparrow}' &= E_{2\uparrow \uparrow} - 8 v_A.
\end{split}
\end{equation}

Applying the forth order perturbation theory gives us:
\begin{equation}
\begin{split}
	E_{\uparrow \uparrow} &= E_{0\uparrow \uparrow} - \frac{18 t^2}{E_{1\uparrow \uparrow}-E_{0\uparrow \uparrow}} \\
	&- \frac{9 t^4}{(E_{1\uparrow \uparrow}-E_{0\uparrow \uparrow})^2}(\frac{1}{E_{2\uparrow \uparrow} - E_{0\uparrow \uparrow}} + \frac{1}{E_{2\uparrow \uparrow}' - E_{0\uparrow \uparrow}}).	
\end{split}
\end{equation}

So the energy difference between parallel and anti-parallel $e_g$ spins of V atoms is:
\begin{equation}
\begin{split}
	E_{\uparrow \uparrow} - E_{\uparrow \downarrow} &= 24 v_A + \frac{9t^4}{(\Delta E_{10})^2}(\frac{1}{\Delta E_{20}} - \frac{1}{\Delta E_{20} - 8 v_A}) \\
	&\approx 24 v_A - \frac{72 t^4 v_A}{(\Delta E_{10})^2(\Delta E_{20})^2}
\end{split}
\end{equation}
where $\Delta E_{10} = E_{1\uparrow \downarrow} - E_{0\uparrow \downarrow} = E_{1\uparrow \uparrow} - E_{0\uparrow \uparrow}$, $\Delta E_{20} = E_{2\uparrow \downarrow} - E_{0\uparrow \downarrow} = E_{2\uparrow \uparrow} - E_{0\uparrow \uparrow}$. 

\end{document}